# Modelling intracellular signalling networks using behaviour-based systems and the blackboard architecture

PEDRO PABLO GONZÁLEZ PÉREZ
CARLOS GERSHENSON
MAURA CÁRDENAS GARCÍA
JAIME LAGUNEZ OTERO
Instituto de Química
Universidad Nacional Autónoma de México
Ciudad Universitaria, 04510, MÉXICO, D.F.
ppgp@servidor.unam.mx  http://132.248.11.4

*Abstract*: This paper proposes to model the intracellular signalling networks using a fusion of behaviour-based systems and the blackboard architecture. In virtue of this fusion, the model developed by us, which has been named Cellulat, allows to take account two essential aspects of the intracellular signalling networks: (1) the cognitive capabilities of certain types of networks' components and (2) the high level of spatial organization of these networks. A simple example of modelling of $Ca^{2+}$ signalling pathways using Cellulat is presented here. An intracellular signalling virtual laboratory is being developed from Cellulat.

## 1 Introduction

Each cell in a multicellular organism receives specific combinations of chemical signals generated by other cells or by themselves. The final effect of the signals received by a cell can be translated in the regulation of the cell metabolism, in the alteration or maintenance of its differentiation state, in the beginning of cellular division and in its death. Once the extracellular signals bind to the receptors, different process of signalling are activated, which generate complex networks of intracellular signalling.

The more experimental data about cellular functioning we have, the more useful the computational models are to understand these signalling processes. The computational models allow the visualization of the components that integrate the network, in addition to being able to predict or to suggest the effect of disturbances caused on determined component or sections of the signalling pathway. A great variety of information processing computational models have been inspired by the functioning of biological systems. Examples of these are the artificial neural networks, genetic algorithms and cellular automata. On the other hand, several information processing systems have constituted different useful approaches to model the functioning of diverse biological systems [16].

Within computer sciences, the artificial intelligence area has constituted one of the main scenarios to model biological systems. This fact responds to the great variety of models, techniques and methods that support this research area, many of which are inherited of disciplines such as psychology, cognitive sciences and neuroscience. Between the main techniques of artificial intelligence and computer sciences commonly used to model cellular signalling networks are artificial neural networks [1 and 18], boolean networks [4], petri nets [7], rule-based systems [3], cellular automata [4], and multi-agent systems [5, 17 and 19].

The high complexity level of the intracellular communication networks causes them to be difficult to model when is considered some of these techniques isolated. Nevertheless, when integrating the most relevant and necessary elements of these techniques in a single computational system, then it should be possible to obtain a model much more robust of the intracellular signalling pathway, and therefore, a better visualization, understanding and prediction of the processes and components that integrate the networks.

The theory of behaviour-based systems must constitute a good approach to the intracellular signalling networks. However, a powerful intracellular signalling model should be achieved when the communication between agents takes place through a shared data structure, in which other cellular structures and elements related with signalling pathways can be explicitly represented. In this sense, the blackboard architecture

[15] results widely appropriate.

In this paper, we postulate that an effective and robust model of intracellular signalling can be obtained when the main structural and functional characteristics of behaviour-based systems and the blackboard architecture are joined. That is, a cell can be seen as a society of autonomous agents, where each agent communicates with the others through the creation or modification of signals on a shared data structure, named "blackboard". The autonomous agents model determinated functional components of the intracellular signalling pathways, such as signalling proteins, enzymes and other mechanisms. The blackboard levels represent different cellular structures committed with the signalling pathway, whereas the different objects created on the blackboard represent secondary messengers, activation or inhibition signals or others elements belonging to intracellular medium.

In this way, when the autonomous agents are used in an intracellular signalling model, the cognitive capabilities of certain components of the signalling pathway can be taken in count; whereas the use of blackboard architecture allows to capture the high level of spatial organization exhibited by the intracellular signalling networks.

This paper is structured as follows: the next section presents an overview of the information processes of signalling intracellular networks. In section 3 some questions about behaviour-based systems are discussed. Section 4 describes the basic components of blackboard architecture and their functioning. In section 5 the intracellular signalling model proposed by us is presented and discussed. In section 6 an example of modelling the $Ca^{2+}$ signalling pathway using the proposed model is shown. Some comments about current and future work and conclusions are made in section 7.

## 2 An overview of signalling intracellular networks

Each cell receives a great number of chemical signals generated by other cells or by themselves. These regulate their metabolism, determine their differentiation, and indicate when to divide or to start a death process. In general the external signals are transmitted to the interior of the cells through membrane receptors activating diverse transduction pathways, which can follow a same way and to generate a final answer or to branch themselves to give rise to others. For example activation of most membrane-associated hormone receptors generates a diffusible intracellular signal called a second messenger. Many different receptors generate the same second messenger. Some intracellular messengers are: cAMP (cyclic AMP), cGMP (cyclic GMP), I3P (Inositol triphosphate), DG (diacylglycerol) and $Ca^{2+}$ (calcium).

The second messenger activation triggers multiple signal transduction pathways like: the adenilate cyclase, the hosphoinositide-calcium, the Mitogen activated protein kinase (MAPK), the JAK/STAT, the sphingomyelin-ceramide, and the ionic channel function receptors pathways.

In particular if we take the $Ca^{2+}$ signalling pathway, the pathway looks like an interconnected network, because the answer is to different stimuli, but the different pathways cross and generate alternative trajectories not easily visualized. The field of signal transduction has centred on the discovery of increasing crosstalk among signalling pathways. Indeed, our understanding of the topology of signalling is continuously evolving. Early work in the signalling field featured the discovery of branched linear pathways, which, resembled trees with a cell surface receptor at the root.

## 3 Behaviour-based systems

The theory of behaviour-based systems [2] provides a new philosophy for the construction of autonomous agents, inspired by ethology. The goal of behaviour-based systems (BBS) is to provide control to autonomous agents. An autonomous agent interacts directly with its problem domain, it perceives its environment through its sensors and acts on it through its actuators. The autonomous agent environment is commonly a dynamic, complex and unpredictable environment, in which the autonomous agent tries to satisfy a set of goals or motivations, varying in time. An autonomous agent decides by itself how to relate its external and internal inputs with its motor actions, in such way that its goals can be satisfied [13]. Adaptation is one of the desirable characteristics in autonomous agents. An autonomous agent is adaptive if it has abilities that allow him to improve its performance in time.

A cell can be seen as an adaptive autonomous agent or as a society of adaptive autonomous agents, where each agent could exhibit a particular behaviour depending on his cognitive capabilities. As an autonomous agent, the cell perceives its external environment through its surface receptors and acts on this one by means of the generation of new signals, which will be able to affect the behaviour of other cells. The cell's external environment is also dynamic and complex, in which an extensive range of signals and combinations of these can exist. Furthermore, like in an autonomous agent, the cell by itself decides how to relate the received external signals (signalling molecules) to its internal signals (secondary messengers), so that their goals (differentiation, proliferation, survival, among others) can be satisfied. Finally, the cell is able also to adapt to its environment. That is, the cell is constantly adapting its behaviour to the changes that take place in the environment and to the signals perceived.

## 4 Blackboard architecture

The blackboard architecture concept was conceived by artificial intelligence researchers in the 1970's. The goal of this research was to handle the problem of shared information between multiple expert agents in problem solving. The architecture was developed for the first time in the system for language understanding, Hearsay II [15]. Later it has been used in a great variety of problem domains and abstracted in many environments for the construction of systems [10 and 14].

Blackboard architecture is defined in terms of three basic components: (1) a set of independent modules, named knowledge sources, which contain specific knowledge about a problem domain, (2) a shared data structure, named blackboard, through which the knowledge sources communicate to each other, and (3) a control mechanism, which determines the order in which the knowledge sources will operate on the blackboard. Figure 1 shows the blackboard architecture components and the relationships among them.

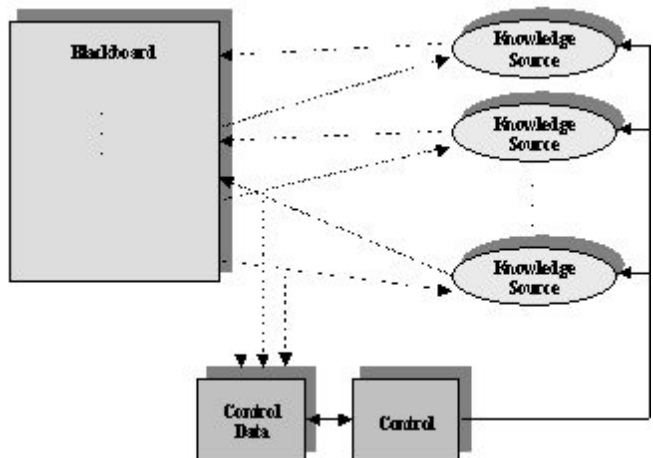

**Fig. 1.** Components of the Blackboard Architecture

The knowledge sources are modular software subsystems that represent different points of view, different strategies and different knowledge types, about how to solve a problem or part of a problem.

The blackboard is used as a central storage for all shared information. The information on a blackboard represents facts and deductions done by the knowledge sources during the problem solving. The knowledge sources produce changes on a blackboard, which incrementally lead to the formation of a solution or acceptable set of solutions for the problem to be solved.

Because the knowledge sources respond opportunistically to changes on the blackboard, a mechanism is necessary that controls these changes and decides, at every moment, which actions should be taken. The control mechanism handles the interaction between the blackboard, the knowledge sources and the outsourcing such as users and control or data acquisition subsystems.

## 5 Cellulat: an intracellular signalling network model

Our proposal consists in modelling the cell as an autonomous agent, which in turn is composed by a society of autonomous agents, which communicates through blackboard with others. That is, in the proposed model the knowledge sources of the blackboard architecture are considered as autonomous agents too. The model proposed here constitutes a refinement and adaptation of an action selection mechanism structured on a blackboard architecture previously developed by us named Internal Behaviours Network (IBeNet) [8, 9, 11 and 12]. Although the IBeNet was initially built to action selection in autonomous agents (physical robots, animats, or artificial creatures simulated on a computer), it constitutes a working environment for the bottom-up modelling of information processing systems characterized by: (1) coordination and opportunistic integration of several tasks in real time, (2) use of several abstraction or context levels for the different types of information that participate in the processing network, (3) decision making, (4) action selection and (5) adaptation.

Although, the intracellular signalling model proposed can be seen as the assignation of a new semantic to the components of the IBeNet, we assume here that this fact is equivalent to the assignation of this same semantic to the components of the blackboard architecture, when the two following considerations are done: (1) knowledge sources are seen as either internal autonomous agents or interface autonomous agents and (2) the blackboard architecture control is distributed now between these two types of autonomous agents. The term "internal autonomous agents" has been used to identify to the autonomous agents which tasks are related with the creation or modification of signals on the blackboard. An internal autonomous agent gets a signal or combination or signals from a determinated blackboard level and transduces these in other signal on the same or other blackboard level. The way in which a signal is transduced depends of the cognitive capabilities of the internal autonomous agent On the other hand, the function of an interface autonomous agent is to establish the communication between the blackboard and the external medium (they are similar to sensors or actuators in BBS). Not all external signals or combinations of these are recognized by an interface autonomous agent, this recognition depends both of the signal characteristics and the cognitive capabilities of the interface autonomous agent.

In Figure 2 the architecture of intracellular signalling model can be appreciated. This is a translation of blackboard architecture shown in Figure 1 taking account the considerations (1) and (2). In this way, it is not necessary to explain the structural and functional details of IBeNet here to understand the intracellular signalling model proposed. It is only necessary to bear in mind the blackboard architecture functioning explained in section 3 and the new considerations mentioned above. The intracellular signalling model has been named Cellulat (a kind of animat which behaves as a cell).

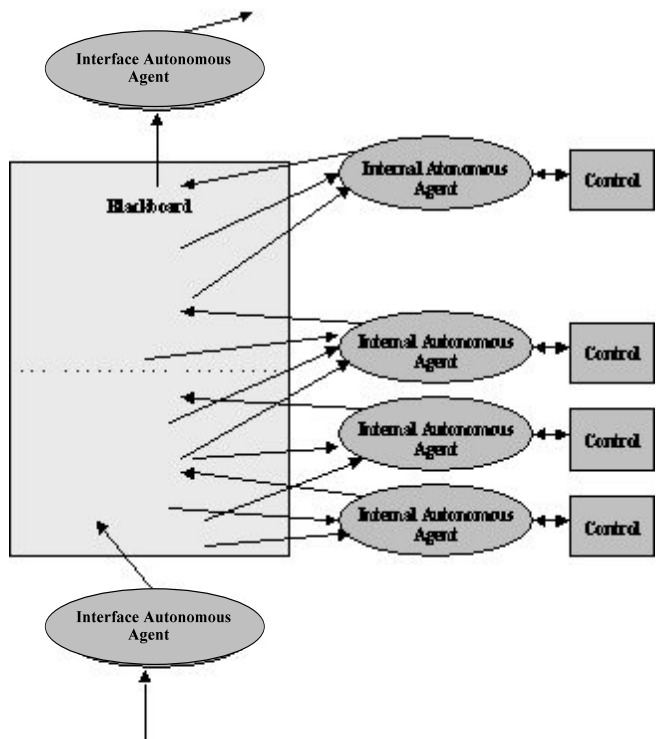

**Fig. 2.** Architecture of the Cellulat

Three main components define the Cellulat structure: the blackboard, the internal autonomous agents and interface autonomous agents.

The blackboard represents the cell's internal medium. The blackboard levels correspond to different cellular structures through which occur the signal transduction. In this way, the cellular membrane, the citosol and the nucleus could be represented as different blackboard levels. The solution elements recorded on the blackboard represent two main types of intracellular signals: secondary messenger molecules and activation/inactivation signals. Both types of signals are synthesized or created by internal autonomous agents and these, either directly or indirectly, promote the activation/inactivation of other internal autonomous agents. Other types of cellular elements or structures can be represented on the blackboard too.

The internal autonomous agents model components of the intracellular signalling network such as proteins, enzymes and other mechanisms necessary to carry out the signal transduction. On the other hand, the interface autonomous agents model the cell surface receptors and the mechanisms for the secretion of signalling molecules. Each agent, independently of his type, has a condition part and an action part; the way in which both parts are linked depends on the complexity of the intracellular component modeled by agent. For that reason, agents who model complex components could use more advanced techniques, such as neural networks, genetic algorithms, or any combination of other techniques, to link both parts. Agents which model less complex components could use more sophisticated but useful techniques, such as production rules, boolean networks or others. The work of both types of agents is event-directed. This is, each intracellular signal registered on the blackboard or each extracellular signal perceived constitutes an event, which could activate or inactivate one or more autonomous agents. When an internal autonomous agent is activated then this one executes his action, which consists in the synthesis or modification of a signal on the blackboard.

**5.1 Cognitive capabilities of Cellulat's agents**

It is known that certain types of proteins exhibit several cognitive capabilities such as pattern recognition, memory and handling fuzzy data [5]. The evolution, learning and emergence of properties have also been suggested in proteins networks [18]. In Cellulat, both types of autonomous agents exhibit several cognitive capabilities including patter recognition, handling uncertainty and fuzzy data, adaptive action selection, memory and learning; which allow to the autonomous agents to exhibit an adaptive behaviour. These cognitive capabilities are supported by several artificial intelligence paradigms including the following approaches: rule-based reasoning, probabilistic reasoning, artificial neural networks, genetic algorithms, boolean networks, and fuzzy logic systems.

### 5.2 Modelling spatial organization in Cellulat

Another important aspect to consider in the modelling of intracellular signalling networks is their spatial organization. Experimental data recently obtained suggest clearly that many intracellular signalling networks exhibit a high level of spatial organization. Intracellular signalling models developed recently approach to this question [6]. Cellulat allows to model the spatial organization taking account two organizational criteria of the blackboard architecture. One is the horizontal organization, given by the different abstraction levels of the blackboard, which allow an intralevel signal processing. The other is the vertical organization, given by columns that cross vertically different blackboard levels. These columns arise as result of the adjoining work of several internal autonomous agents that operate at a same section of blackboard, which cover different blackboard levels. We have named these columns "agency columns". Convergence and divergence of agency columns could occur, and these processes could be related with evolution and learning of the signalling network [11]. In this way, the model proposed allows to model intracellular signalling pathways taking account their spatial organization. That is, the two information processing levels present in Cellulat (horizontal and vertical) allow to establish a "topology preserving map".

## 6 Modelling the Ca2+ signalling pathway using Cellulat: A first approach

Figure 3 illustrates a first approach of the intracellular signalling pathway modelling using Cellulat. Ins this case, we have modelled the $Ca^{2+}$ signalling pathway to a still heavy resolution level. This is, this first model not yet reveals structural details of the signalling network components.

As can be appreciated in Figure 3, the interface autonomous agents model the G-protein-linked surface receptors, whereas the internal autonomous agents model components of the signalling network such as G protein, phospholipase C-beta, protein kinase C, mechanism to release calcium, and others proteins and enzymes. On the blackboard can be created different types of signals by the proteins, enzymes and others mechanisms that operate on it. Examples of these signals are the followings: PI-biphosphate, inositol triphosphate, dyaclyglycerol, $Ca^{2+}$, and others signals indicating the activation state of the proteins and enzymes. In this representation, the blackboard levels have not been identified.

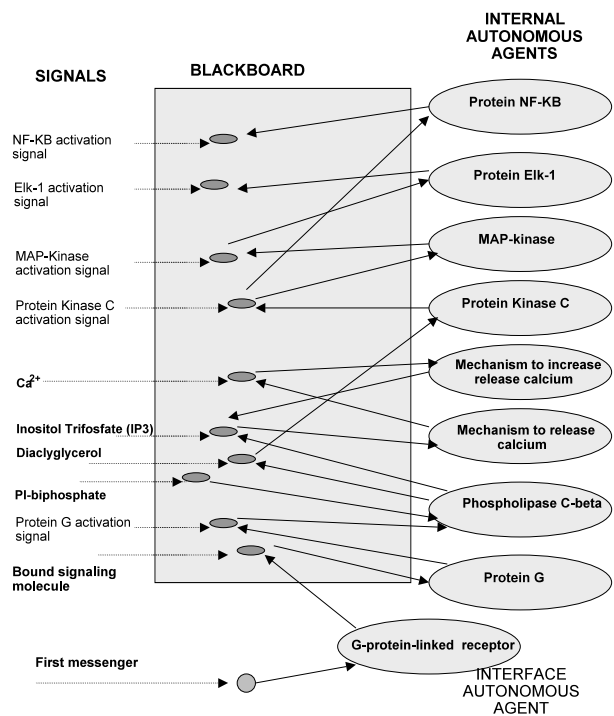

**Fig. 3.** Cellulat model of the Ca2+ signalling pathway.

In the model shown in Figure 3 it is necessary to consider the following viewpoint: Since an internal or interface autonomous agent can be structured by other autonomous agents, then an internal autonomous agent representing to a protein kinase C (or to others proteins or enzymes) could encapsulate many proteins kinase C, where each one of these is an internal autonomous agent too. In this way, components of a same type could jointly be encapsulated, which simplifies comprehension.

## 7 Conclusions

In this paper we have discussed why the fusion of the behaviour-based systems theory and the blackboard architecture constitutes a very suitable approach for the modelling of cellular signalling networks, from a model developed by us which has been named Cellulat. As a result of this fusion, Cellulat allows to model two basic aspects of the cellular signalling networks: (1) the cognitive capabilities of certain types of proteins and enzymes, and (2) the high level of spatial organization fo the signalling networks. (1) was obtained when considering the adaptive autonomous agents as functional components of the model, and using different techniques of artificial intelligence to support the cognitive capabilities mentioned in Section 5.1; whereas (2) was provided by the shared data structure, named blackboard, which allows to model horizontal, vertical and spatial information processing.

A simple example of modelling of $Ca^{2+}$ signalling pathways using Cellulat was shown. Although still much work to develop is left, as it is the case of the modelling of the signalling network components at a fine resolution level, we believe the proposed model constitutes a new and useful approach for the modelling and understanding of intracellular signalling networks.

Presently, Cellulat is being used in the modelling of the signal transduction related with cellular senescence. Our future work is directed towards the creation of a cellular signalling virtual laboratory, from the developed computational model. This virtual laboratory must allow to effect lesions on certain components or sections of the signalling pathways, and to visualise the obtained cellular behaviour as consequences of such lesions; helping to understand how the different cellular signalling pathways interact.